# Non-Ohmic negative longitudinal magnetoresistance in two-dimensional electron gas

Yang-Yang Lyu[1,2], Xian-Jing Zhou[3], Zhi-Li Xiao[1,4,*], Roxanna Fotovat[1,4], Jing Xu[3], Gobind Basnet[4], Yong-Lei Wang[2,*], Dafei Jin[3], Ralu Divan[3], Hua-Bing Wang[2], and Wai-Kwong Kwok[1]

[1]*Materials Science Division, Argonne National Laboratory, Argonne, Illinois 60439, USA*

[2]*Research Institute of Superconductor Electronics, School of Electronic Science and Engineering, Nanjing University, Nanjing 210093, China*

[3]*Center for Nanoscale Materials, Argonne National Laboratory, Argonne, Illinois 60439, USA*

[4]*Department of Physics, Northern Illinois University, DeKalb, Illinois 60115, USA*

*Correspondence to: xiao@anl.gov; yongleiwang@nju.edu.cn

Negative longitudinal magnetoresistance (NLMR) has been reported in a variety of materials and has attracted extensive attention as an electrotransport hallmark of topological Weyl semimetals. However, its origin is still under debate. Here, we demonstrate that the NLMR in a two-dimensional electron gas can be influenced by the measurement current. While the NLMR persists up to 130 K, its magnitude and magnetic field response become dependent on the applied current below 60 K. The tunable NLMR at low and high currents can be best attributed to quantum interference and disorder scattering effects, respectively. This work uncovers non-Ohmic NLMR in a non-Weyl material and highlights potential effects of the measurement current in elucidating electrotransport phenomena. We also demonstrate that NLMRs can be a valuable phenomenon in revealing the origins of other properties, such as negative MRs in perpendicular magnetic fields.

## I. INTRODUCTION

Certain materials placed in a magnetic field exhibit magnetoresistance (MR), where their resistance changes with magnetic field [1]. MRs in magnetic materials have been extensively explored and widely used in applications [2] while intriguing MR phenomena such as extremely large magnetoresistance [3] and linear magnetoresistance [4] have been observed recently in non-magnetic systems. Typically, MRs are measured with magnetic fields perpendicular to the current [3,4]. On the other hand, MRs in other field orientations are also of great interest, as they reveal new effects such as anisotropic magnetoresistance in magnetic materials [5]. In particular, the recent discovery of topological Weyl semimetals has drawn attention to the behavior of longitudinal magnetoresistances (LMRs) when the magnetic field is parallel to the current (**H**//**I**) [6]. Here, Weyl nodes with opposite chiralities exist in pairs and in the presence of parallel magnetic and electric fields, a chiral anomaly manifests itself as an intervalley pumping of electrons between the Weyl nodes with opposite chiralities. The electron pumping generates charge imbalance between the two nodes resulting in an electric current that can only be balanced by intervalley scattering. In clean samples, the intervalley scattering time is long. Thus, a Weyl semimetal under **H**//**I** is expected to exhibit non-Ohmic negative LMRs (NLMRs) [7,8], with potential applications in nonlinear electronics and as a platform for the development of a topological Fermi-liquid theory [9].

The focus on NLMR which has been observed in numerous materials [6,8-23] in recent years follows the prediction that it may be one of the few electrotransport hallmarks of Weyl semimetals exhibiting chiral anomalies. Alternative origins without invoking chiral anomaly have also been proposed for NLMRs in materials with trivial [23-25] or topological [14,15,19,20,25,26] states. For example, NLMRs were found in the disordered topological insulator $TlBi_{0.15}Sb_{0.85}Te_2$ [19] and in quantum wells [23], where they were attributed to magnetic-field sensitive percolation effect [19] and distorted current paths [23,27], respectively. On the other hand, the non-Ohmic behavior of NLMRs has been hardly investigated in experiments, except for the non-linear current-voltage characteristics in $Bi_{0.96}Sb_{0.04}$, which was attributed to a charge pumping effect induced by the chiral anomaly [9].

Here we report on non-Ohmic NLMRs in a non-Weyl material. They reveal a new mechanism for the non-Ohmic behavior of the NLMR in addition to the chiral anomaly. More importantly, the behavior of NLMRs is found to depend on the amplitude of the measurement current. Magnetotransport measurements were conducted on a two-dimensional electron gas (2DEG) in a GaAs heterostructure and NLMRs were observed at temperatures up to 130 K. At $T < 60$ K, the value of the NLMR changes when the applied current is varied. Furthermore, the measurement current also dictates the behavior of the resistance versus magnetic field $R(H)$ curve. Quantitative analysis of the $R(H)$ relationship, coupled with the change in the NLMR's dimensional nature from 3D to quasi-2D with increasing current, reveal that quantum interference is most likely the dominating origin for the NLMRs at low currents while their behavior at high currents can be best attributed to disorder related effects. These results relate the non-Ohmic behavior in the NLMR to current-induced delocalization. They highlight the role of the measurement current on the observation and interpretation of electrotransport phenomena. We also demonstrate that NLMRs can be used to reveal the origins of other properties, such as the negative MRs in perpendicular magnetic fields.

## II. EXPERIMENTAL DETAILS

The measured samples are custom-made Hall sensors composed of GaAs/GaAlAs heterostructures whose layer structure is presented in Fig.1(a). The GaAs/GaAlAs heterostructure was patterned with photolithography into Hall bars with width of $L_y = 17$ μm and voltage lead distance of $L_x =$ 40 μm. Ohmic contacts to the 2DEG were fabricated by diffusing Au. An image of the Hall bar samples is shown in Fig.1(b).

We conducted DC resistance measurements in magnetic fields up to 9 Tesla using a custom-built magnetotransport system. Angle dependence of the resistance was obtained by placing the sample on a precision, stepper-controlled rotator (Attocube model ANR 51) with an angular resolution of 0.05°. We measured the magnetoresistance of Hall bars co-located on the same wafer-chip and share the same applied current (see Fig.1(b)). In order to verify the homogeneity of the sample we used two pairs of voltage leads (A&B and C&D) to measure the resistance $R$ along the current flow direction. The resistance $R$ from both pairs of voltage leads behave the same. Here we focus on results from voltage leads A&B, with additional data from voltage leads C&D presented in the supplement (Fig.S1) [28]. The traverse resistances $R_{xy}$ was measured using voltage leads A and E. We define the magnetoresistance as $MR = (R - R_0)/R_0$, where $R$ and $R_0$ are resistivities at a fixed temperature with and without magnetic field, respectively.

**III. RESULTS AND DISCUSSION**

From the temperature dependence of the zero-field resistance $R_0(T)$ and the Hall resistance $R_{xy}(T)$ with an out-of-plane magnetic field of $H = 9$ T (see inset of Fig.1(c)), we obtained the electron density and mobility and present them in Fig.1(c). The electron density is nearly temperature-independent, with a value of $8.71 \times 10^{15}/m^2$ at $T = 3$ K. On the other hand, the mobility decreases with increasing temperature, changing from $\mu = 1.77$ m$^2$/Vs at $T = 3$ K to $\mu = 0.72$ m$^2$/Vs at $T = 150$ K. $R(H)$ and $R_{xy}(H)$ curves obtained at $T = 3$K and $I = 1$ μA in out-of-plane magnetic fields (**H**⊥**I**) in Fig.1(d) show the Shubnikov de Haas oscillations and quantum Hall effect plateaus, respectively.

Figure 2(a) and Fig.S2 present data on the $MR(H)$ curves and the corresponding $R(H)$ curves at $T = 3$ K and at various angles between the magnetic field and the current (see inset of Fig.2(a) for the definition of angle $\theta$). Negative MRs can be clearly seen at $\theta = 0°$, i.e., **H**//**I** and quickly disappear when the magnetic field is tilted away from the direction of the current, e.g., at $\theta \geq 5°$. The occurrence of the negative MR limited to near $\theta = 0°$ and the shape of the $MR(H)$ curve are akin to those observed for NLMRs attributed to chiral anomaly [6,8,9,12,13]. Figure 3(a) shows $MR(H)$ curves obtained at $\theta = 0°$ and at various temperatures. The NLMRs in our sample also diminish with increasing temperature and disappears at $T \sim 130$ K, similar to the temperature behavior found in Weyl semimetals.

The new discovery is the dependence of NLMRs on the magnitude of the applied measurement current as demonstrated by the $R(H)$ curves presented in Fig.2(b) for $\theta = 0°$ and $T = 3$ K. For $I > 40$ μA, the shape of the $R(H)$ curve does not change significantly, although the amplitude of the NLMR increases with current. Below $I < 30$ μA, however, clear changes in the shape of the $R(H)$ curves can be identified. Figure 3(b) presents $MR(H)$ curves at $\theta = 0°$ and $I = 1$ μA at temperatures up to $T = 160$ K. Comparing them to those in Fig.3(a) for $I = 100$ μA, we see clear differences in both their shapes and temperature behavior at $T < 60$ K while they look similar for $T \geq 60$ K. This unusual temperature dependence of the MR to the applied current is unambiguously displayed in Fig.4 for $I = 1$ μA and $I = 100$ μA at $H = 9$ T, where the two $MR(T)$ curves overlap for $T \geq 60$ K and diverge below $T < 60$ K. Furthermore, above $T > 60$K, the two curves overlap at all magnetic fields as shown in Fig.S3 for data obtained at $T = 70$ K and 100 K, although the absolute values of the resistance are larger at the higher measurement current. These results evidently reveal non-Ohmic NLMRs in our samples and potential effects of the measurement current on the observation and understanding of NLMRs.

Currently, non-Ohmic behavior of NLMRs has been reported only in the Weyl semimetal $Bi_{0.96}Sb_{0.04}$ [9] where the chiral-anomaly induced nonlinear longitudinal electrical conductivity $\sigma_L$ is found to exhibit a monotonically quadratic dependence on the electrical field $E$, i.e. $\sigma_L \sim (HE)^2$, where $H$ is the applied magnetic field along the current direction [9]. Figure 5(a) presents the measured current dependence of the longitudinal magnetoresistance $R(I)$ at $T = 3$ K for $0$ T $\leq H \leq$ 9 T and Fig.S4 displays the derived $\sigma_L$ versus $E$ curves for a few representative magnetic fields. The green line in Fig.S4 for $H = 6$ T shows that $\sigma_L(E)$ follows a quadratic behavior at small $E$ values ($E < 0.35$ V/mm). With increasing $E$, however, $\sigma_L$ reaches a maximum at a magnetic-field dependent $E$ value and then decreases with further increase of $E$. A similar behavior of $\sigma_L(E)$ is also observed at $H = 0$ T. This behavior differs from that of a Weyl semimetal where $\sigma_L$ is expected to be independent of $E$ in the absence of an external magnetic field.

Non-Ohmic behavior could arise from Joule heating which increases the electron temperature [29-33]. As presented in the inset of Fig.1(c), the resistance obtained at $I = 1$ µA increases monotonically with temperature. Thus, Joule heating would induce larger resistance with increasing current at $I > 1$ µA. This differs from the results displayed in Fig.5(a) for $T = 3$ K and various magnetic fields, which show non-monotonical $R(I)$ relationship with a resistance minimum occurring at a field-dependent current value that increases slightly from $I_m \sim 30$ µA for $H = 0$ T to $I_m \sim 37$ µA for $H = 9$ T. Consequently, Joule heating is not the cause of non-Ohmic NLMRs at $I < I_m$ while it cannot be excluded for $I > I_m$. Data in Figure 5(b) for $H = 0$ T and various temperatures indicate that this non-monotonical behavior exists over a wide range of temperature ($T \leq 20$ K). At high temperatures ($T > 20$ K), the sample resistance does increase monotonically with increasing current. As demonstrated by the results for $T = 70$ K and 100 K in Fig.S3, however, MR is independent of the measurement current, although the resistance is non-Ohmic. Since the MR does change with temperature at $T > 60$ K (see Fig.4), the current-independent MRs suggest that the monotonical non-Ohmic behavior at high temperatures is not dominated by Joule heating. Besides Joule heating, the electric field $E$ induced by the measurement current can cause deviations of the electron distribution function from the Fermi-Dirac form, which is considered to be the origin for the non-Ohmic resistance of two-dimensional electron gas in crossed electric and out-of-plane magnetic fields ($\mathbf{H} \perp \mathbf{I}$) [30]. The non-monotonical $R(I)$ relationship in our sample at low temperatures ($T < 20$ K) as presented in Fig.5 looks very similar to that reported in Ref.30 for out-of-plane magnetic fields. It may not be a surprise because the resistance shows a similar $R(I)$ relationship in both the presence and the absence of a magnetic field, as revealed by data in Fig.5(a). That is, deviations of the electron distribution from the Fermi-Dirac function can be contributors to the non-Ohmic behavior in the NLMRs observed in our samples. However, theoretical analyses of the $R(I)$ relationship in the presence of a magnetic field indicate that electric-field induced changes to the electron distribution function suppress the resistance monotonically with increasing current and cannot account for the enhancement in the resistance at high currents [30].

In a low dimensional system such as our 2DEG samples, quantum interference effects can play an important role in magnetotransport properties. For example, weak localization is one of the quantum interference effects that is considered to be the origin of the negative MRs in graphenes [34] and quantum wells [35,36] under crossed electrical and magnetic fields. When localization occurs in the presence of strong and dilute scatterers [37], the probability for an electron to return to the same impurity is enhanced by the magnetic field, subsequently reducing its probability for an electron to scatter off other impurities and thereby resulting in a negative MR [36]. Furthermore, non-Ohmic behavior could occur in a 2D system with weak localizations at high dc currents that lead to $\delta > a$, where $\delta \sim E$ is the drift distance along the $E$ direction and $a$ is the radius of the scatterers. Such a current-induced 'delocalization' leads to a linear increase of the resistance at high currents as observed in GaAs quantum wells [36]. Our sample at low currents and low magnetic fields (see Fig.2(b) and Fig.S5) show typical negative MRs from weak localization, i.e., $MR \sim -H^{1/2}$ [23]. Thus, the enhancement in the resistance by high currents (Fig.5) in our 2DEG system can also originate from current-induced 'delocalization'. In fact, the $R(I)$ relationship at high currents in Fig.5(a) is similar to that observed in GaAs quantum wells [36], regardless of the alignment between the electric and magnetic fields.

The occurrence of current-induced 'delocalization' can also account for the change in the shape of the $R(H)$ curves at different currents (Fig.2(b)). As discussed above, at low currents the $R(H)$ relationship, particularly the NLMRs at low magnetic fields, can be attributed to weak localization, although further investigations are needed to understand the non-monotonical $R(H)$ behavior including the positive MRs at high magnetic fields. When electrons are 'delocalized' at high currents, the shape of the $R(H)$ curves, as demonstrated in Fig.6 for $I = 100$ μA and $T = 3$ K, exhibit a parabolic $R(H) \sim -H^2$ behavior over a wide range of magnetic fields, in contrast to the $R(H) \sim -H^{1/2}$ behavior at low currents where weak localization dominates.

A $R(H) \sim -H^2$ relationship for **H//I** is theoretically predicted for topological insulators [26] and can also roughly describe the magnetic field dependence of NLMRs arising from chiral anomaly, i.e., $\sigma_L \sim H^2$ (see Fig.6 for a comparison). Besides the deviations of the parabolic fit from experimental data at high fields (Fig.6), our samples are neither 3D topological insulators nor Weyl semimetals. On the contrary, our results can be consistently explained with a model proposed by Herfort et al. [38], which successfully accounts for the negative MR in quantum wells for an in-plane magnetic field that is oriented perpendicular to the applied current, **H⊥I**. By considering the magnetic field dependent reduction of an additional contribution to impurity scattering between the plane-wave states in the 2DEG and localized states in the fluctuation potential in the *d*-doped layer, they obtained the following magnetic field dependence of the sample resistance:

$$R = (R_0 - R_{min}) \exp\left[-aH^2/(1+bH) - \ln(1+bH)/2\right] + R_{min} \qquad (1)$$

where $R_0$ is the zero-field resistance, $R_{min}$ is the lowest resistance, $a = 2\pi^2 e^2 \Delta s^2 \lambda^2 / h^2$, $b = 2\pi e q \pi \lambda^2 / h$. Here $\Delta s$ is the separation of the 2DEG and the $\delta$-doped layer, $\lambda$ is the correlation length of the fluctuation potential, $q$ is a dimensionless parameter, and $e$ and $h$ are the electron charge and the Planck constant, respectively.

As demonstrated in Fig.6 for *T* = 3 K, the *R*(*H*) curves at *I* = 100 µA can be described by Eq.1. The obtained characteristic lengths $\Delta s \lambda$ are presented in the inset of Fig.6, whose values at low temperatures are consistent with those reported by Herfort et al. [38]. Since the electron density *n* (see Fig.1(c)) changes by less than 3% over the entire temperature range, we could not verify the reported relationship between *n* and $\Delta s \lambda$. On the other hand, the temperature dependence of $\Delta s \lambda$ in the inset of Fig.6 and the electron mobility in Fig.1(c) reveal a clear association between them. As the model predicts, our samples also exhibit negative MRs for in-plane magnetic fields but at **H**⊥**I** (see Fig.S5). Since the model considers the effects of impurity scattering on the plane-wave states in the 2DEG, it can naturally explains the 2D nature of the NLMRs presented in Fig.2(a) for *I* = 100 µA. *R*(*H*) curves in Fig.S6 for out-of-plane magnetic fields ($\theta$ = 90°) show pure positive and negative MRs at *I* > 50 µA and *I* < 15 µA, respectively, revealing the 2D and 3D nature of the mechanisms responsible for the negative MRs at high and low currents, respectively. This suggests again a current-driven change in the origin of NLMRs. However, the shape of the *R*(*H*) curves in Fig. S6 at small currents, e.g., at *I* = 1 µA, is clearly different from the NLMR ($\theta$ = 0°) in Fig.2(b), indicating potential additional contribution from other mechanisms such as memory effects [36,39] that occur only in perpendicular magnetic fields. In other words, NLMRs may be used to identify or distinguish the origin of the negative MRs in out-of-plane fields.

## IV. CONCLUSION

In summary, we observed non-Ohmic negative longitudinal magnetoresistances (NLMRs) up to 130K in a 2DEG heterostructure, GaAs/GaAlAs. We found that the values of NLMRs depend on the amplitude of the applied current for $T < 60$ K. We attribute the observed non-Ohmic behavior in the NLMR to current-induced delocalization, with the dominance of quantum interference and disorder related effects on the NLMRs at low and high currents, respectively. These results reveal the existence of possible new mechanisms for the observed non-Ohmic NLMR, in addition to the chiral anomaly that has been widely applied to understand NLMRs and related phenomena. We also show that NLMRs can be helpful in uncovering the origins of other properties, such as the negative MRs in perpendicular magnetic fields.

**Acknowledgements**

Magnetotransport measurements were supported by the U.S. Department of Energy, Office of Science, Basic Energy Sciences, Materials Sciences and Engineering. R. F. and Z. L. X. acknowledge supports by the National Science Foundation under Grant No. DMR-1901843. Use of the Center for Nanoscale Materials, an Office of Science user facility, was supported by the U.S. Department of Energy, Office of Science, Office of Basic Energy Sciences, under Contract No. DE-AC02-06CH11357. Y. Y. L, Y. L. W. and H. B. W. acknowledge supports by the National Natural Science Foundation of China (61771235 and 61727805).

**Figure Captions**

**Fig. 1.** Characteristics of the samples. (a) Schematic (not-to-scale) of the GaAs/AlGaAs heterostructure. (b) Optical image of the GaAs Hall bar sample. Current flows between electrical contacts $I^+$ and $I^-$. Two pairs of voltage leads (A&B and C&D) were used to measure the resistance $R$ along the current flow direction and one pair of voltage leads (A&E) was used to determine the transverse resistance $R_{xy}$. Besides a proportional decrease due to the smaller separation, the resistance $R$ obtained between leads C&D behaves exactly the same as that between leads A&B, indicating the homogeneity of the sample. All resistance data reported in this article are from leads A&B. (c) Temperature dependence of the electron density and mobility. The electron density is calculated from the $R_{xy}(T)$ curve obtained in an out-of-plane magnetic field of $H = 9$ T, which allows us to derive the mobility from the zero-field resistance $R_0(T)$. Inset of (c) presents the associated $R_{xy}(T)$ and $R_0(T)$ curves. The data were taken at $I = 1\mu A$. (d) $R(H)$ and $R_{xy}(H)$ curves obtained at $T = 3K$ and $I = 1$ μA in out-of-plane magnetic fields (**H**⊥**I**), showing Shubnikov de Haas oscillations and quantum Hall effect plateaus, respectively.

**Fig. 2.** Current-dependent negative longitudinal magnetoresistance. (a) Magnetoresistance $MR(H)$ curves at $I = 100$ μA and various magnetic field orientations, where $MR = [R(H) - R_0]/R_0$ and $R_0$ is the sample resistance at $H = 0$ T. The corresponding $R(H)$ curves and data obtained at more field orientations are presented in Fig.S2. The inset presents a schematic of the angle $\theta$ between the magnetic field **H** and the direction of the current **I**, with $\theta = 0°$ for **H**//**I** and $\theta = 90°$ for **H**⊥**I**. Negative magnetoresistances occur when the magnetic field is aligned very close to the direction of the current and disappear at $\theta \geq 5°$. (b) Longitudinal magnetoresistance $R(H)$ curves at measurement currents of 1 μA ~ 100 μA. Both the amplitude of the MR and the shape of the $R(H)$ curve depend on the value of the measurement current. All the data were taken at $T = 3$ K.

**Fig. 3.** Temperature dependence of the longitudinal magnetoresistance for different currents. (a) $MR(H)$ curves at $I = 100$ µA and various temperatures, showing a monotonical temperature-dependence. Negative MR behavior remains nearly constant for $T < 60$ K, decreases monotonically with increasing temperature, and varnishes at $T \approx 130$ K. (b) $MR(H)$ curves at $I = 1$ µA and various temperatures, indicating a non-monotonical temperature-dependence. For $T \geq 60$ K, the MRs are exactly the same as those at $I = 100$ µA, i.e. gradually diminishing with increasing temperature. For $T < 60$ K, the negative MR becomes more pronounced at higher temperatures. The shape of the $R(H)$ curve is also temperature-dependent, differing from those obtained at $I = 100$ µA in (a). All the data were taken at $\theta = 0°$.

**Fig. 4.** Comparison of the temperature dependence of the longitudinal magnetoresistances at different measurement currents. The data were taken from the $MR(H)$ curves in Fig.3. Results for $H = 9$ T at $I = 1$ µA and 100 µA are used as an example to highlight the current-dependent MR temperature regime ($T < 60$ K).

**Fig. 5.** Current dependence of the longitudinal magnetoresistance. (a) Data obtained at $T = 3$ K and various magnetic fields. The $R$ versus $I$ relationship shows a non-monotonical behavior. The dip occurs at a field-dependent current value, which increases slightly from $I_m \sim 30$ µA for $H = 0$ T to $I_m \sim 37$ µA for $H = 9$ T. (b) Results obtained at $H = 0$ T and various temperatures. The $R$ versus $I$ relationship shows a non-monotonical behavior at $T \leq 20$ K and becomes monotonical at $T > 20$ K. The maximum of the normalized resistance versus temperature relationship at 100 µA appears at $T \sim 35$ K. All the data were taken at $\theta = 0°$.

**Fig. 6.** Analysis of the NLMR using various models. Experimental data (symbols) are obtained at $T$ = 3 K, $I$ = 100 μA and **H**//**I**. Dashed blue line represents a parabolic relationship $R(H) = R_0(1 - \alpha H^2)$ with $\alpha$ = 3.956×10$^{-4}$ T$^{-2}$. Solid green line is derived from the longitudinal magnetoconductivity of Weyl fermions $\sigma_L = \sigma_0(1 + \beta H^2)$ with $\sigma_0 = 1/R_0$ and $\beta$ = 3.9×10$^{-4}$ T$^{-2}$. Solid red line is a fit from Eq.1 (see text). In all three cases, $R_0$ = 960.62 Ω is the measured resistance at zero field. Inset presents the temperature dependence of the determined characteristic length $\Delta s\lambda$ by fitting the experimental $R(H)$ curves using Eq.1.

**Figure 1**

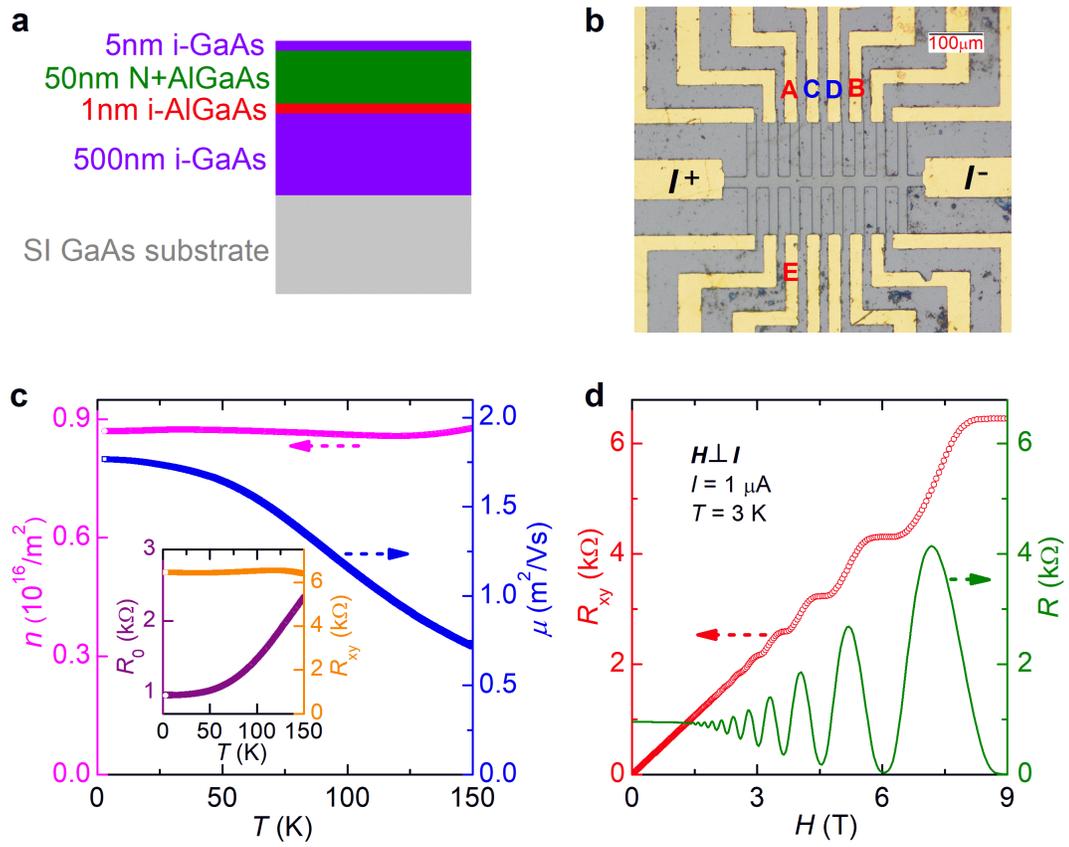

**Figure 2**

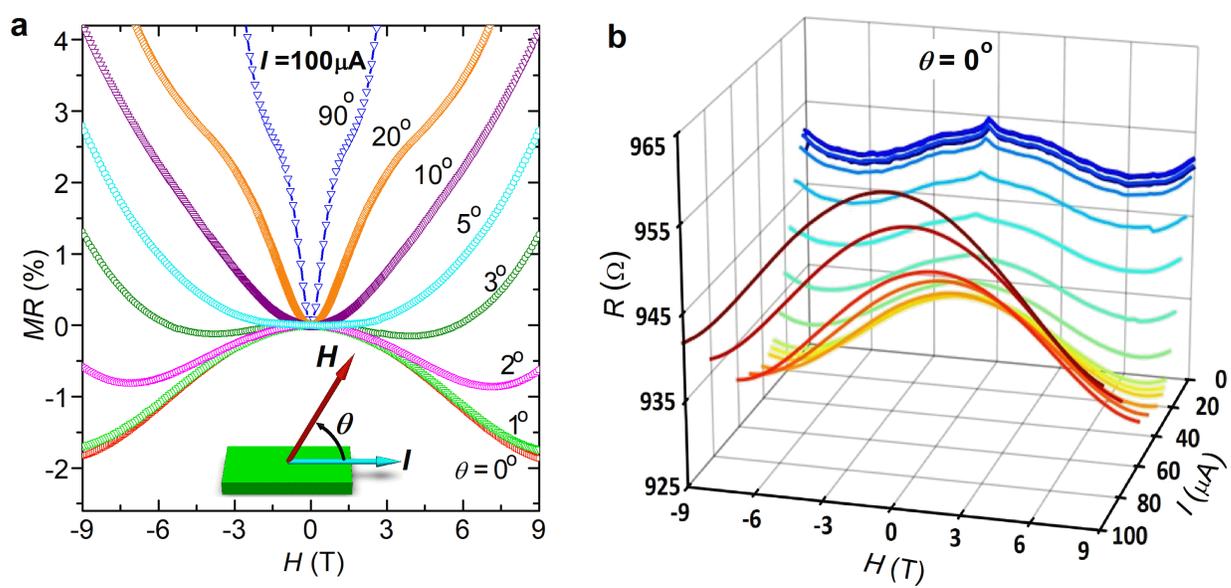

**Figure 3**

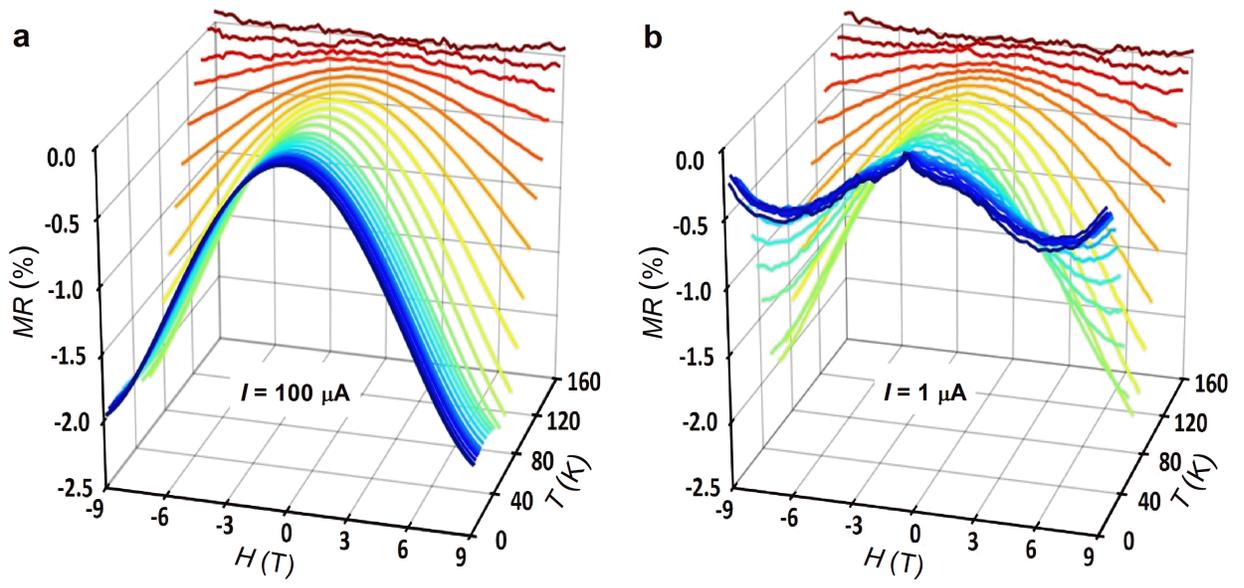

**Figure 4**

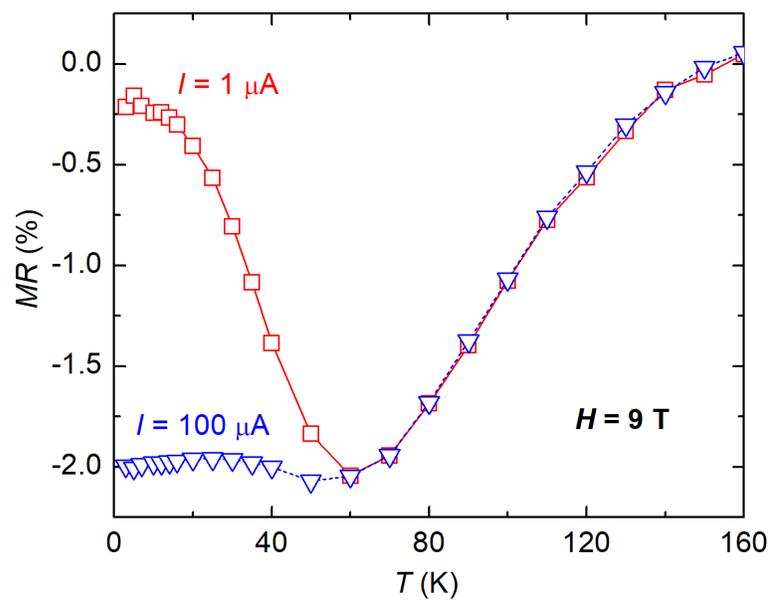

**Figure 5**

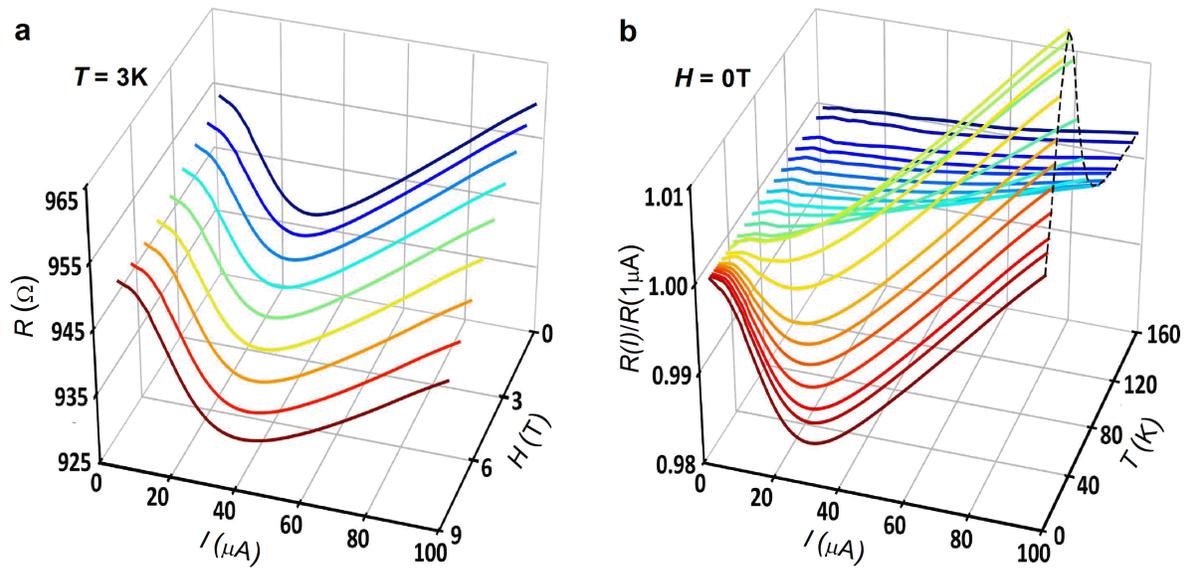

**Figure 6**

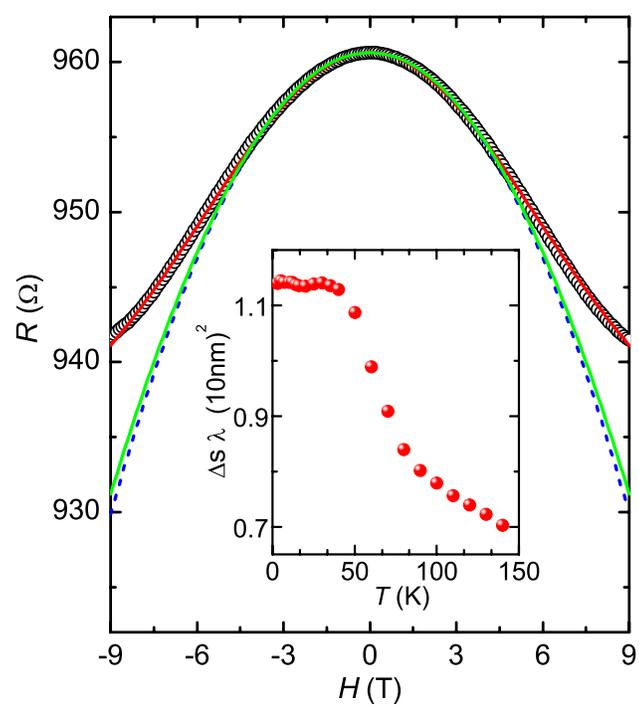